# Controlling the dimensionality of low-Rm MHD turbulence experimentally

Nathaniel T. Baker · Alban Pothérat · Laurent Davoust · François Debray · Rico Klein

**Abstract** This paper introduces an experimental apparatus, which drives turbulence electrically in a liquid metal pervaded by a high magnetic field. Unlike past magnetohydrodynamic (MHD) setups involving a shallow confinement, the experiment presented here drives turbulence whose dimensionality can be set anywhere between three-dimensional and quasi two-dimensional. In particular, we show that the dimensionality and componentality of the turbulence thus generated are in fact completely fixed by the single parameter $l_z(l_i)/h$, which quantifies the competition between the solenoidal component of the Lorentz force and inertia acting on a turbulent structure of the size of the forcing scale $l_i$. This parameter is fully tunable thanks to the three operating settings at hand: the injection scale, the intensity of the electric forcing and the magnitude of the magnetic field. Thanks to the very high number of measuring probes and fast acquisition rate implemented in this experiment, it is possible to reliably measure the finest features of the inertial range on a scale-wise basis.

N. T. Baker · A. Pothérat · R. Klein
Coventry University, AMRC
Priory Street, Coventry CV15FB, UK

N. T. Baker · L. Davoust
Grenoble-INP/CNRS/Univ. Grenoble-Alpes
SIMaP EPM, F-38000 Grenoble, France

N. T. Baker · F. Debray
LNCMI-EMFL-CNRS, UGA, INSA, UPS
25 Ave. des Martyrs, 38000 Grenoble, France

## 1 Introduction

The present paper introduces an experimental apparatus capable of fixing the dimensionality of turbulence anywhere between three-dimensional (3D) and quasi two-dimensional (quasi-2D). 3D and 2D turbulence are known to display radically opposite dynamics (see for instance [1] for a thorough review of the subject), where the former features a direct cascade of energy from the injection scale down to the small dissipative scales controlled by viscosity, while the latter features an inverse cascade of energy from the injection scale up to the large structures controlled by the geometry of the system. However, very little is known about the dynamics of turbulence, which simultaneously possesses 2D and 3D turbulent scales. The word "dimensionality" itself can take several meanings and may be understood either as a synonym for velocity gradients in the bulk, or for the number of non-zero components of the velocity field. Talking about the dimensionality of turbulence is all the more ambiguous, as turbulent flows are customarily described on a scale-wise basis. There is thus a need for experimental tools capable of delivering reliable statistics over a wide range of turbulent structures.

As of today, a class of experimental and numerical studies has emerged, which deals with flows combining wide ranges of interacting 2D and 3D turbulent structures, involving velocity fields with either two or three components [2], [3], [4], [5]. Although there are several ways to force some two-dimensionality at the laboratory scale (for instance by geometrical confinement, application of a background rotation or magnetic field), quantifying the dimensionality of the resulting flow proves to be a difficult task. Celani *et al.* and Xia *et al.* proposed to measure the dimensionality of their flows through the aspect ratio of the height of the flow to the forcing scale. Though intuitive, this approach returns a relatively crude estimate, which is based on the geometry of the system, rather than the flow itself.

The experiment presented in this paper takes place within the low-Rm magnetohydrodynamic (MHD) framework. Historically speaking, MHD turbulence has often been used to investigate the dynamics of 2D turbulence [6], [7], [8], [9], [10]. It has also recently been employed to investigate 3D turbulence [11], [5]. The significant advantage of MHD at the laboratory scale to tackle the problem of turbulence dimensionality comes from the existence of a clean theoretical prediction for a cutoff lengthscale separating 2D and 3D turbulent structures, when evolving in a wall-bounded domain [12]. More specifically, turbulent structures larger than this cutoff lengthscale are quasi-2D, whereas turbulent structures smaller than this cutoff lengthscale are 3D. As we will show, our experiment exploits this very scaling law to precisely control the dimensionality of turbulence.

The present paper splits down into five main parts. We will start by briefly reviewing the theory on which this experiment relies in section 2, before detailing the experimental setup in section 3. Section 4 will give the experimental procedure used to study forced statistically steady MHD turbulence, while section 5 will validate the precision at which such turbulence can be measured, by assessing the convergence of the statistics. We will close this paper with section 6, in which we present some global features of the flow generated in this experiment, and illustrate how its dimensionality can be precisely controlled.

## 2 Theoretical background

Depending on authors, the term "dimensionality" may actually refer to two distinct concepts. On the one hand, it can be understood as a synonym for velocity gradients in the bulk. Owing to this definition, three-dimensionality is thus associated to the spatial dependence of physical quantities with respect to any given spatial coordinate. The two-dimensional limit then refers to a situation where the bulk is fully correlated over a preferential direction (such as the direction of an imposed magnetic field for instance). On the other hand, dimensionality may also be understood as the number of non-zero components of the velocity field. We shall call this latter type of dimensionality "componentality", in order to distinguish it from the former.

When dealing with turbulence, it is customary to think of the flow on a scale-wise basis. Speaking about the dimensionality and/or componentality of turbulence per say then becomes difficult, since some turbulent flows are found to simultaneously feature what could a priori be called 2D and 3D turbulent scales. As a matter of fact, it is possible to observe in nature or laboratory experiments so-called coherent vortices, that is to say long-lived and large-scale vortices, which stand out from a background of otherwise random motions. In shallow configurations, these coherent vortices may be topologically 2D (i.e. spatially invariant along their axis of rotation), and emerge as a result of 2D dynamics (i.e. an inverse energy cascade). And yet, these coherent structures may concurrently exist with topologically and dynamically 3D structures in the background. With this point of view, dimensionality then becomes a function of the size of the turbulent structure at hand.

One of the main features of low-$Rm$ MHD is the production of eddy currents by velocity gradients. The global impact of these eddy currents on MHD flows may be interpreted as a "pseudo-diffusion" of momentum in the direction of the magnetic field [12], which is driven by the solenoidal component of the Lorentz force. The end result of this diffusive process being the two-dimensionalization of the flow over the diffusion length $l_z$ in the direction of the magnetic field. Denoting $\rho$ and $\sigma$ the density and electric conductivity of the fluid respectively, as well as $B_0$ the magnitude of the magnetic field, the eddy currents are drawn on a timescale given by the Joule time $\tau_J = \rho/\sigma B_0^2$. A turbulent structure of width $l_\perp$ is then diffused by the solenoidal component of the Lorentz force over the distance $l_z$ over the timescale $\tau_{2D} = \tau_J (l_z/l_\perp)^2 = (\rho/\sigma B_0^2)(l_z/l_\perp)^2$. Assuming this turbulent structure lays in the inertial range of fully developed MHD turbulence with a velocity $u'(l_\perp)$, the main process opposing its two-dimensionalization is inertia, by means of scale-wise energy transfers occurring over the eddy turnover time $\tau_u = l_\perp/u'(l_\perp)$. The competition between these two processes introduces the following estimate for the diffusion length $l_z(l_\perp)$:

$$\frac{l_z(l_\perp)}{h} = \sqrt{N}\,\frac{l_\perp}{h}, \tag{1}$$

where $N = \sigma B_0^2 l_\perp / \rho\, u'(l_\perp)$ is the local (in scale space) interaction parameter based on the width of the structure in question and its velocity. Equation (1) gives a succinct way of characterizing the dimensionality of the structure by comparing its diffusion length $l_z(l_i)$ to the height $h$ of the channel in which it evolves [5]. Namely, $l_z(l_\perp)/h \ll 1$ implies that the turbulent structure of width $l_\perp$ is topologically 3D, as the Lorentz force is not quick enough to diffuse its momentum across the channel before the structure yields its energy to the cascade. Conversely, $l_z(l_\perp)/h \gg 1$ means that the turbulent structure of width $l_\perp$ is quasi-2D, as the inertial transfers take place over a much longer time scale than that required for the Lorentz force to diffuse its momentum across the experiment. In that sense, the Lorentz force may be seen as a filtering process, which segregates between topologically 2D and 3D scales over the height of the box. The global dimensionality of the flow may then be estimated experimentally by associating it to the diffusion length of the injection scale, whose size $l_i$ is imposed by the forcing mechanism.

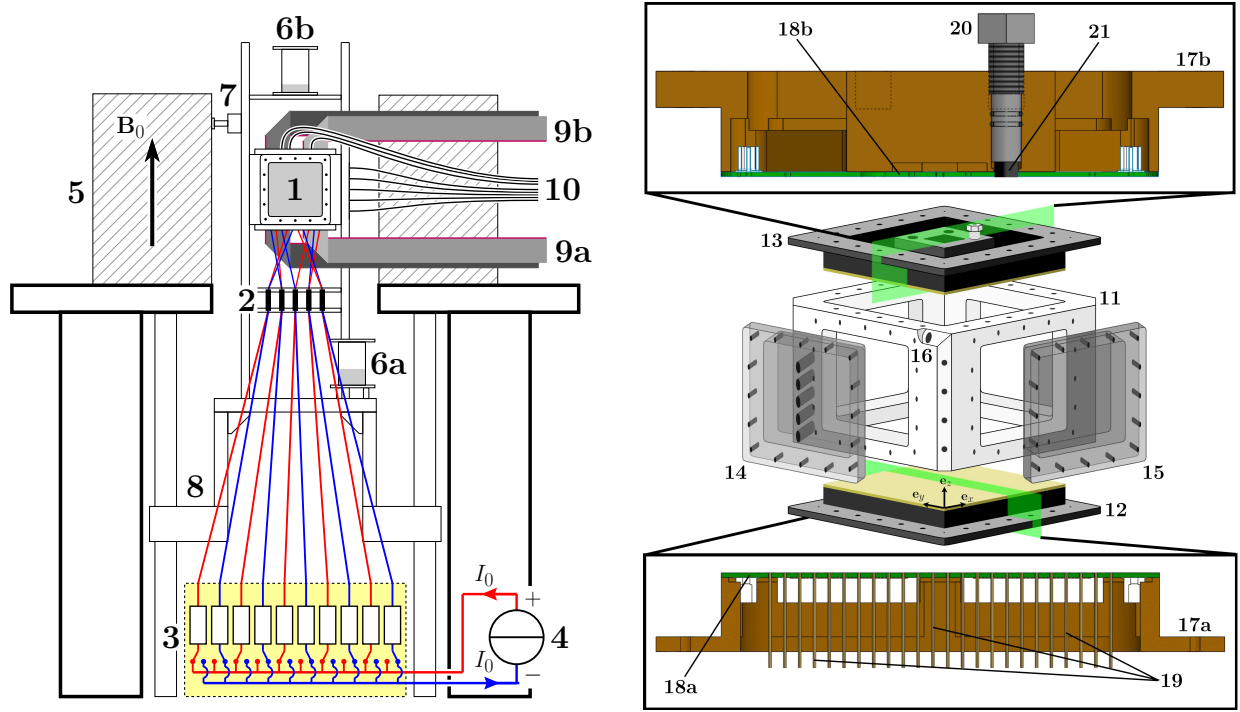

**Fig. 1** Left: General layout of the experimental setup: (1) Vessel filled with Galinstan; (2) Current injection panel; (3) Injection table; (4) Power supply; (5) Superconducting magnet; (6a) Fill tank; (6b) Overflow tank; (7) Alignment adjustment spacer; (8) support platform; (9a) Bottom EPV signals to amplifiers; (9b) Top EPV signals to amplifiers; (10) PUDV signals to DOP4000. Right: exploded view of the vessel. (11) Polyacetal frame; (12) Injection plate (i.e. bottom Hartmann plate); (13) Top Hartmann plate with one ultrasound probe fitted; (14) Ultrasound plate; (15) Polycarbonate window; (16) Galinstan outlet (inlet locate in diagonally opposite corner). Bottom insert: profile view of the injection plate. (17a) Brass frame; (18a) Bottom PCB; (19) Injection electrodes protruding through the plate. Top insert: profile view of the top plate. (17b) Brass frame; (18b) Top PCB; (20) Holding device for the ultrasound probe (machined from a nylon screw); (21) Ultrasound transducer.

## 3 Apparatus design and instrumentation

### 3.1 General layout of the Flowcube

The centerpiece of the experiment presented here is a closed vessel filled with Galinstan, a eutectic alloy of gallium, indium and tin, which is liquid at room temperature. Galinstan is characterized by a density $\rho = 6400\,\mathrm{kg/m^3}$, kinematic viscosity $\nu = 4 \times 10^{-7}\,\mathrm{m^2/s}$ and electric conductivity $\sigma = 3.4 \times 10^6\,\mathrm{S/m}$. Statistically steady turbulence is driven electrically by inducing Lorentz forces, which originate from forcing a total DC current $I_0$ through the vessel, while simultaneously applying a static and uniform magnetic field $B_0$. The forced electric current enters and exits the experiment thanks to a periodic array of electrodes located along the bottom wall. The working principle of the driving mechanism in the Flowcube is very similar to Sommeria's experimental setup [8]. The crucial difference between the two rigs however, is that we consider a thick layer of fluid (100 mm) to allow for 3D effects, whereas [8] considered a thin layer of fluid (20 mm) to observe quasi-2D effects exclusively.

The vessel is mounted around a rectangular parallelepiped frame made of polyacetal, which possesses an inner square base of width $L = 150\,\mathrm{mm}$, and a height $h = 100\,\mathrm{mm}$. The Flowcube is designed as a modular platform whose faces can be independently fitted (cf. figure 1). The top and bottom faces are closed by so-called "Hartmann plates" (cf. section 3.2), which consist of $148\,\mathrm{mm} \times 148\,\mathrm{mm}$ wide printed circuit boards (PCBs) mounted on polyamide coated brass frames. The sides of the vessel are closed by side plates mounted into $100\,\mathrm{mm} \times 100\,\mathrm{mm}$ openings. Up to now, diverse side plates have been developed, each offering a specific functionality:

- a plate entirely meshed by a $38 \times 38$ array of probes, to measure the local electric potential with a spatial resolution of 2.5 mm;
- a plate fitted with five ultrasound transducers enabling ultrasound velocimetry;
- a plate fitted with seven intrusive electric potential probes, enabling the measurement of all three components of the electric potential gradient in the bulk;
- a plate offering the vertical profile of electric potential flush to the side wall (described in Pothérat & Klein [5]).

The modularity of this experiment makes it possible to either change the size of the vessel by building a new frame (cf. [11] and [5] who used a smaller version of the Flowcube, mounted around a 100 mm wide cubic frame), or to build new plates equipped with alternate devices to either drive and/or monitor the flow. For instance, one could very well imagine a new plate, which would drive the flow mechanically with a propeller or an oscillating grid. For the sake of concision, we shall focus in this paper on the side plate fitted with ultrasound transducers exclusively. This particular plate consists of a polycarbonate window, which carries five horizontally aligned ultrasound transducers stacked on top of each other. Taking the altitude reference $z = 0$ mm at the surface of the bottom plate, the five ultrasound transducers are located at heights $z = 12$ mm, $z = 31$ mm, $z = 50$ mm, $z = 69$ mm and $z = 88$ mm respectively (see figure 1.right).

Airtightness of the vessel is ensured by O-rings on the outer side of the frame, complemented by internal silicone seals cast within the gaps between plates. Galinstan is supplied to the main chamber from a fill tank connected to one of the bottom corners. A similar tank connected to the diagonally opposite top corner receives the metal overflow. Before liquid metal is allowed in, a thorough cleansing of the whole experiment is performed: it is vacuumed and flushed with argon five times to remove oxygen. This step is mandatory to delay Galinstan oxidization, as gallium oxides tend to yield very poor electric contacts with either the electrodes or the measuring probes. The filling takes place under an inert argon atmosphere, at about 1 bar pressure.

3.2 The Hartmann plates

The Hartmann plates refer to the plates laying perpendicularly to the magnetic field, as these are the plates along which Hartmann boundary layers develop. Both top and bottom plates are fitted with 484 potential probes each mounted flush to the wall, arranged in a cross pattern (see figure 2). They give access to the electric potential along the top and bottom walls, with a resolution of 2.5 mm. The patterns found on the top and bottom plates are mirror symmetrical, such that each potential probe found on the bottom plate has an exactly overhanging counterpart. In addition, the top plate is fitted with three vertically aligned ultrasound transducers, giving access to three distinct vertical profiles of the vertical velocity component .

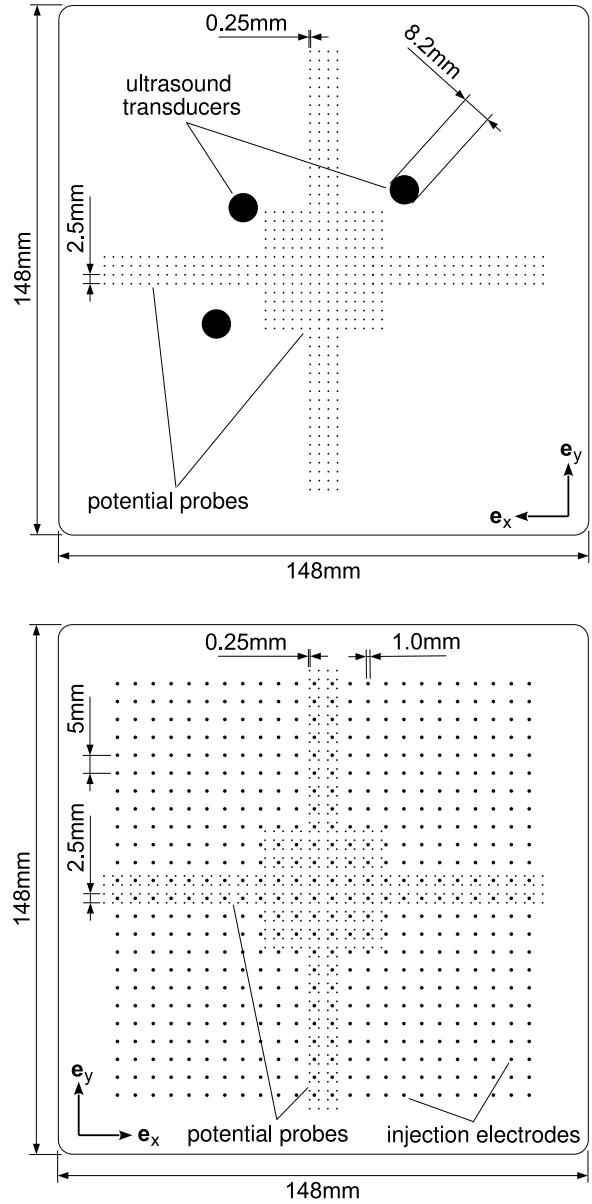

**Fig. 2** Sketch of the Hartmann plates' surfaces in direct contact with Galinstan. Top: top Hartmann plate. Bottom: injection plate.

The heart of each Hartmann plate is a three-layered PCB manufactured by PCB electronics SA, composed of a 1.6 mm thick outer layer of ROGERS 4003C (a high performance hydrocarbon ceramic) pressed with standard FR4 epoxy inner layers. The outer ROGERS layer is in direct contact with Galinstan, while the inner FR4 layers have electric tracks etched in them to extract the potential signals. These signals are picked up by ribbon cables from the edges of the PCB, and then channeled to the acquisition system. Potential probes actually consist of 0.25 mm diameter copper plated vias filled by 0.20 mm diameter copper wires. Each wire was individually soldered on to its track. In addition, the bottom plate (also referred to as the injection plate) is equipped with a $24 \times 24$ array of 1 mm diameter copper electrodes each distant by 5 mm. These electrodes were directly glued onto the PCB. Before fitting the Hartmann plates into the vessel, the electrodes and potential probes were polished and gold plated to improve the initial electric contact between them and Galinstan. The unprecedented number of probes and electrodes introduced in this experiment was chosen to drive and measure turbulence over the widest range of scales as possible, and enable its measurement with sufficient spatial resolution.

3.3 Turbulence driving mechanism

The flow is driven in the Flowcube by applying a static, uniform and vertical magnetic field, while simultaneously forcing DC current through the electrodes of the bottom plate. The forcing mechanism is illustrated in figure 3: the radial component of the current density field $\mathbf{j}$ interacts with the vertical magnetic field $\mathbf{B}_0$ to induce azimuthal Lorentz forces $\mathbf{f} = \mathbf{j} \times \mathbf{B}_0/\rho$. These forces ultimately act as sources of vertical vorticity $\omega_z \mathbf{e}_z$, centered on top of each electrode in use [8].

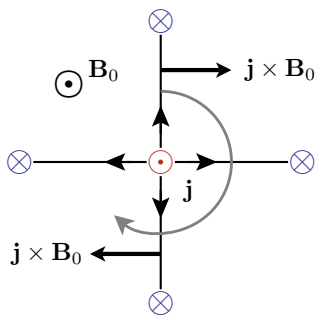

**Fig. 3** Physical principle behind the electric forcing of the flow. The circular arrow depicts the horizontal motion of fluid particles as a result of the azimuthal Lorentz forces.

The experiment was hosted by the high magnetic fields laboratory in Grenoble (LNCMI-G), which granted access to two of their magnets to conduct the experiments. A superconducting magnet with a bore of inner diameter 450 mm was used to deliver magnetic fields up to 4 T, while higher fields (from 5 to 10 T) were accessible thanks to a resistive magnet of inner diameter 376 mm. These two magnets use radically different technologies, both having their pros and cons. For instance, the superconducting magnet gives access to high magnetic field at a relatively low electric cost, since the coil has no electric resitivity. The reduced electric cost necessary to operate a superconducting magnet is however balanced by its heavy consumption of cryogenic fluids to keep the temperature of the coil below 4 K. In both cases, magnetic field inhomogeneities at the vessel's level were of the order of 5% in both magnets.

DC electric current is supplied to the Flowcube by an EA-PSI 9080-300 DC power supply manufactured by Elektro-Automatik GmbH & Co.KG (up to 300 A in total, ca. 7.5 kW, with an output ripple below 100 mA). The interfacing between the power supply and the injection plate is provided by the injection table, consisting of $2\,\Omega \pm 0.25\%$ resistors, each mounted in series between the power supply and one electrode (cf. figure 1.left). These high precision resistors ensure that the total incoming current is evenly split among the electrodes, thus preventing any forcing dissymmetry during the experiments. Indeed, the contact resistance between copper and Galinstan is so low (typically about $1\,\mathrm{m}\Omega$), that any perturbation (such as the temporary presence of gallium oxide, or boundary effects along the edges of the forcing area) may completely off balance the overall current distribution should there not be a larger resistance in the circuit. The injection table used in this experiment enabled the simultaneous connection of up to 100 electrodes among the 576 available on the bottom Hartmann plate.

Two different injection patterns were used in the following experiments (cf. figure 4). First, the case $l_i = 5$ mm refers to a 50 mm wide injection patch formed by an array of $10 \times 10$ electrodes each distant by $l_i = 5$ mm. In this configuration, the patch is focused in the center of the injection plate. Second, the case $l_i = 15$ mm refers to a 110 mm wide injection patch formed by an array of $8 \times 8$ electrodes each distant by $l_i = 15$ mm. In this disposition, the injection pattern spans the whole width of the vessel. In both cases, the electrodes were alternately connected to the positive and negative poles of the power supply, thus driving a square periodic array of counter-rotating vortices.

Turbulence occurs when the array of vortices destabilizes under the action of inertia. In the general case,

perimental and theoretical studies have however been carried out to clarify the destabilization of exclusively quasi-2D arrays [8], [13]. In particular, they have shown that in such a configuration, the transition to turbulence followed the pairing of same-sign vortices. In a channel bounded by two horizontal Hartmann walls, the transition turns out to occur when the non-dimensional parameter $Rh = (h\,I_0/B_0\,\sigma\nu)^{1/2}/2L$ exceeds the critical threshold $Rh = 1.78$. Physically speaking, this implies that past a given point, the Hartmann friction exerted by the boundary layers on the bulk (which is the prominent dissipating mechanism in the quasi-2D limit) is not strong enough to outweigh inertia. In all the experimental runs presented hereafter, the parameter $Rh$ was set above the aforementioned threshold.

### 3.4 Pulsed Ultrasound Velocimetry

The Flowcube is equipped with eight identical ultrasound transducers, consisting of electrically insulating and non-magnetic epoxy piezoelectric crystals of diameter $d = 8\,\text{mm}$, and emitting frequency $f_e = 8\text{MHz}$. The entire data acquisition chain for Pulsed Ultrasound Doppler Velocimetry (PUDV) was handled by the DOP-4000 unit, based on the original design by Willemetz [14]. Five of the ultrasound probes are stacked on top of each other, yielding the $u_x$ velocity profile along the $x$ direction. The other three transducers are mounted on the top plate, giving access to the profile of vertical velocity at three different locations. The probes are mounted flush to the side wall, in direct contact with Galinstan.

The ultrasound signals were generated and recorded using the DOP4000 manufactured by Signal-Processing SA, to which four channels may be connected at once. Depending on the required dynamic range (i.e. on the amplitude of the velocity to measure), the sampling frequency typically ranged between $0.5\,\text{Hz}$ for the lowest velocities (found along the vertical direction) and $10\,\text{Hz}$ for the highest horizontal velocities. Ultrasound measurements require the flow to be seeded with tracing particles to reflect acoustic waves. Using Galinstan has this one advantage that it inevitably introduces gallium oxides, which are effective acoustic reflectors [15].

PUDV is a measuring method originating from the medical field, where non-invasive measurements of blood flows are required. It is based on the principle of echography, which analyses the echoes resulting from the reflection of sound waves upon seeding particles. The method implemented in the Flowcube is known as the "pulse-echo" method, in which the transducer subsequently acts as a an emitter by generating a short wave packet, then as a receiver by recording the resulting the turbulence driven in the Flowcube is a combination of 3D and quasi-2D turbulent structures, for which the destabilization mechanisms are still open to debate and far beyond the scope of this paper (cf. for instance [11] for a brief description of the different steady and unsteady regimes observable in the Flowcube). Some ex-echo. The velocity profile is deduced from the phase shift existing between two consecutive echoes [14], [16].

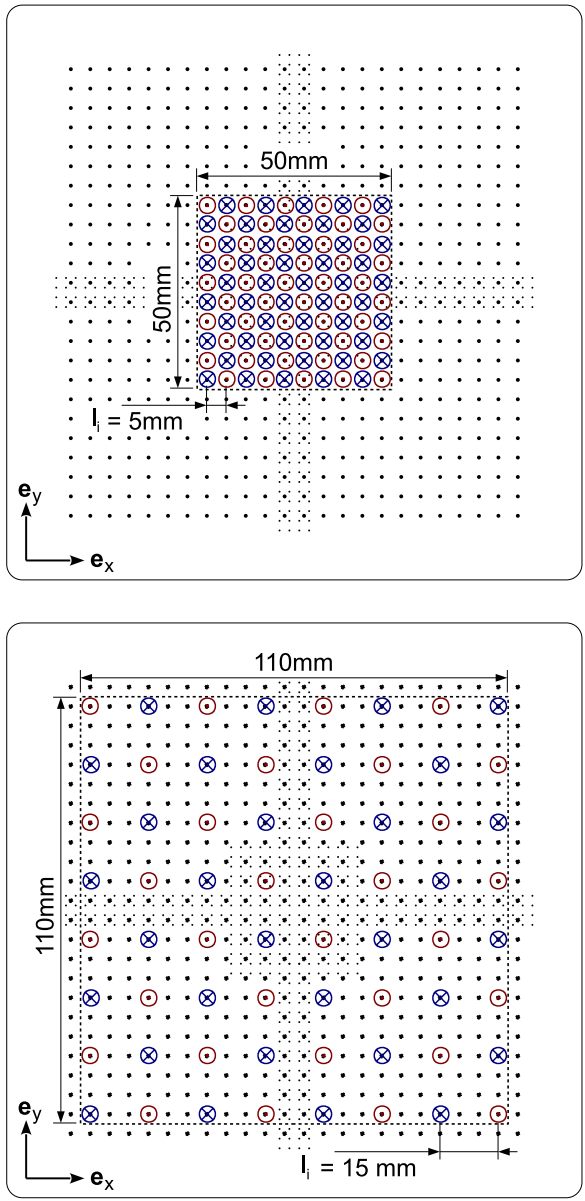

**Fig. 4** Sketch of the two different injection patterns used. Top: $10 \times 10$ array of electrodes used in the configuration $l_i = 5\,\text{mm}$. Bottom: $8 \times 8$ array of electrodes used in the configuration $l_i = 15\,\text{mm}$. The polarity of the electrodes in use is indicated by symbols: $\odot$ for a positive pole; $\otimes$ for a negative pole. The dashed contours delimit the forcing patch, i.e. the region of space where turbulence is actually sustained.

3.5 Electric Potential Velocimetry

Electric potential velocimetry is a robust technique that has been extensively used throughout the years in liquid metal experiments [8], [17], [10], [11]. This particular method of measurement is extremely valuable in liquid metal MHD, as the electric potential measured along no-slip walls laying perpendicularly to the magnetic field happens to be strongly linked to the velocity field. Indeed, Kljukin & Thess [18] showed that in the $Ha \gg 1$ and $N \gg 1$ limit (where $Ha = B_0 h\sqrt{\sigma/\rho\nu}$ is the Hartmann number), measuring the electric potential at the wall is equivalent to measuring the stream function right outside the Hartmann layer. These two quantities are linked according to

$$\psi = -\frac{\phi}{B_0}. \qquad (2)$$

In other words, measuring the electric potential field at a fine enough resolution can yield the two-component velocity field right above and below the bottom and top Hartmann walls via

$$\mathbf{u}_\perp = \nabla \times \left[ \psi(x,y)\,\mathbf{e}_z \right]. \qquad (3)$$

The derivatives appearing in the above relationship are evaluated experimentally by considering a square formed by four adjacent potential probes M, N, P and Q (cf. figure 5), at which the electric potential is known. The center of this square is called O. As mentioned earlier, the potential probes are uniformly spaced by $d = 2.5\,\mathrm{mm}$. The square's diagonal $l$ is thus given by $l = 3.5\,\mathrm{mm}$. Both $x$ and $y$ components of the velocity field ($u_x = \mathbf{u}_\perp \cdot \mathbf{e}_x$ and $u_y = \mathbf{u}_\perp \cdot \mathbf{e}_y$ respectively) may be determined with second order precision at the fictitious point O, by using the following finite difference schemes:

$$u_x(O) = \frac{\phi(N) - \phi(M) - \phi(P) - \phi(Q)}{\sqrt{2}\,B_0\,l} + o(l^2) \qquad (4)$$

and

$$u_y(O) = \frac{\phi(N) + \phi(M) - \phi(P) - \phi(Q)}{\sqrt{2}\,B_0\,l} + o(l^2). \qquad (5)$$

The electric potential signals were recorded in our experiment with an operational amplifiers pack built by neuroConn GmbH. This pack enabled the simultaneous measurement of 767 analog channels, which were directly connected to the potential probes. For all experiments, the sampling frequency was set to $f_s = 250\,\mathrm{Hz}$, and the data was recorded with 24bit precision. The highest gain was used, which gave a $\pm 170\,\mathrm{mV}$ dynamic range (depending on the operating conditions, the amplitude of the electric potential at the Hartmann walls was typically in the range $50\,\mu\mathrm{V}$ to $5 \times 10^4\,\mu\mathrm{V}$). The amplifiers were left to warm up for at least 10 mn before any measurement was performed to obtain drift-free signals.

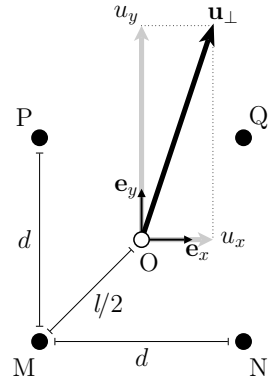

**Fig. 5** Calculation of the two-component velocity field $\mathbf{u}_\perp$ at fictitious point O, using electric potential signals $\phi$ at points M, N, P and Q.

The ambient noise level picked by the amplifiers pack was monitored at the beginning and end of each run. As it turns out, the worst signal to noise ratio was found to be $\mathrm{SNR_{dB}} = 28\,\mathrm{dB}$ (which corresponds to a noise to signal amplitude ratio of 4%), while the best signal to noise ratio was $\mathrm{SNR_{dB}} = 57\,\mathrm{dB}$ (which corresponds to a noise to signal amplitude ratio of 0.1%).

## 4 Experimental protocol and range of parameters

4.1 Range of parameters

The Flowcube offers three distinct control parameters to fix the properties of the flow: the injection scale $l_i$ (defined as the distance separating two adjacently connected electrodes), the magnitude of the magnetic field $B_0$ (measured by the Hartmann number $Ha = B_0 h\sqrt{\sigma/\rho\nu}$), and the total electric forcing $I_0$ (measured by the Reynolds number $Re^0 = I_0/\pi\nu N_e\sqrt{\sigma\rho\nu}$, where $N_e$ is the total number of connected electrodes [19]).

As Pothérat and Klein [5] showed however, the resulting turbulence is in fact fully characterized by only two parameters: the intensity of the turbulent fluctuations (measured by the Reynolds number $Re = u'_\mathrm{bot} h/\nu$) and the dimensionality of the injection scale $l_z(l_i)/h$. In the above, $u'_\mathrm{bot}$ is defined as the rms value of the turbulent fluctuation measured along the bottom plate, thus is representative of the turbulent kinetic energy injected into the system.

Tables 1 and 2 report typical values of these different parameters for the two injection patterns $l_i = 5$ mm and $l_i = 15$ mm, with a total electric forcing $Re^0 = 25590$ in both cases (corresponding to $N_e = 100$, $I_0 = 300$ A on the one hand, and $N_e = 64$, $I_0 = 192$ A on the other). It is interesting to note that despite $Re^0$ being kept constant in both cases, $Re$ unequivocally increases with $Ha$, which reflects the increase of the intensity of turbulent fluctuations with the magnitude of the applied magnetic field [20].

**Table 1** Range of non-dimensional parameters for the case $l_i = 5$ mm and $Re^0 = 25590$ ($N_e = 100$, $I_0 = 300$ A).

| $B_0$ [T]         | 1     | 3     | 5     | 7     | 10    |
|-------------------|-------|-------|-------|-------|-------|
| $u'_{\rm bot}$ [m/s] | 0.180 | 0.230 | 0.240 | 0.250 | 0.270 |
| $Ha$              | 3644  | 10930 | 18220 | 25510 | 36440 |
| $Re$              | 44000 | 58000 | 60000 | 64000 | 67000 |
| $l_z(l_i)/h$      | 0.23  | 0.59  | 0.97  | 1.3   | 1.7   |

**Table 2** Range of non-dimensional parameters for the case $l_i = 15$ mm and $Re^0 = 25590$ ($N_e = 64$, $I_0 = 192$ A).

| $B_0$ [T]         | 1     | 3     | 5     | 7     | 10    |
|-------------------|-------|-------|-------|-------|-------|
| $u'_{\rm bot}$ [m/s] | 0.130 | 0.180 | 0.200 | 0.230 | 0.250 |
| $Ha$              | 3644  | 10930 | 18220 | 25510 | 36440 |
| $Re$              | 32000 | 45000 | 50000 | 57000 | 62000 |
| $l_z(l_i)/h$      | 1.3   | 3.4   | 5.3   | 6.9   | 9.4   |

4.2 Experimental procedure

The space of parameters $(I_0, B_0)$ was scanned for the two injection patterns described earlier. The total current $I_0$ was adjusted such that the current flowing through one electrode ranged from 1 to 7 A in steps of 1 A ($Re^0 \in [4264, 29850]$), while $B_0$ was set within the range 0.25 to 10 T ($Ha \in [911, 36440]$).

The present study focuses on the dynamics of statistically steady turbulence. The main objective was therefore to log sufficiently long data series in order to get meaningful statistics, which required turbulence to be steadily forced. Throughout the recording, the total injected current was carefully monitored to ensure no electrical asymmetries existed. A typical experimental run consisted of the following steps:

1. apply the magnetic field $B_0$, wait for the magnetic field to be stabilized (a few seconds);
2. record electric potential offsets for 3 mn;
3. turn power supply on at set point $I_0$, wait for the flow to reach statistical steadiness;
4. launch acquisition of electric potentials: 18 mn long series in total, split in 6 distinct files of 3 mn each;
5. launch acquisition of ultrasound transducers: 10000 profiles subsequently recorded using the horizontal transducers located at heights $z = 12, 50$ and $88$ mm, as well as a vertical transducer;
6. turn power supply off, wait for the flow to decay;
7. record electric potential offsets for 3 mn and find faulty channels (if any) by comparing to offsets measured at the beginning.

5 Statistical convergence

Investigation of the dynamics of turbulence requires the computation of second and third order moments. An essential validation step to assert that the Flowcube can indeed drive and analyze turbulence of controlled dimensionality is thus to evaluate the degree of precision to which the aforementioned quantities can be obtained. In scale space, they are known as the second and third order structure functions of the turbulent velocity increment $\delta \mathbf{u}'(\mathbf{x}, \mathbf{r}) = \mathbf{u}'(\mathbf{x}+\mathbf{r}) - \mathbf{u}'(\mathbf{x})$, given by $\langle \|\delta \mathbf{u}'\|^2 \rangle(\mathbf{r})$ and $\langle \|\delta \mathbf{u}'\|^2 \delta \mathbf{u}' \rangle(\mathbf{r})$ respectively. For the sake of the present argument, the operator $\langle \cdot \rangle$ represents an ensemble average. Practically speaking, it is calculated under the assumptions of ergodicity and homogeneity by combining temporal and spatial averages respectively. The second order structure function is linked to the scale wise distribution of turbulent kinetic energy, while the third order structure function is related to the scale-by-scale energy transfers [1]. Although the former is not known for posing convergence problems, computation of the latter turns out to be trickier. Mathematically speaking, this behavior comes from the fact that the probability density function (PDF) of the random variable $\delta \mathbf{u}'$ is centered around zero. As such, any odd moment (such as the third) is a signed quantity prone to cancellations during the averaging operation. The calculation process therefore requires a large amount of samples to reach an accurate and reliable estimate, whose exact number is however delicate to assess ahead of time.

Fortunately, Podesta et al. [21] introduced a procedure which quantifies how accurate the estimate for the $n^{th}$ moment is, given a number of independent observations of a random variable. Conversely, this procedure can also be used to predict, from a limited population of samples, how many observations are required to achieve a given accuracy. Let us briefly review the procedure below using a generic random variable $X$. By definition, the $n^{th}$ order moment of the random variable $X$ is given by

$$\langle X^n \rangle = \int_{-\infty}^{\infty} X^n f(X) \, dX, \tag{6}$$

where $f(X)$ is the probability density function associated to the random variable $X$, and $\langle \cdot \rangle$ should be understood as an ensemble average. In practice, $\langle X^n \rangle$ is approximated experimentally by the estimate $M_n$, which is found by averaging over a finite number of realizations. By definition,

$$M_n(N_s) = \frac{1}{N_s} \sum_{i=1}^{N_s} X_i^n, \tag{7}$$

where $N_s$ refers to the number of independent observations of $X^n$, and $X_i^n$ refers to the $i^{th}$ occurrence of $X^n$. It is only in the limit $N_s \to \infty$ that $\langle X^n \rangle$ and $M_n(N_s)$ are rigorously equal. $M_n$ is itself a random variable, whose PDF depends on $N_s$. The shape of this PDF may then be characterized by a mean $\mu_n(N_s)$ and a standard deviation $\sigma_n(N_s)$.

Podesta et al.'s procedure relies on the observation that the ratio $\sigma_n(N_s)/\mu_n(N_s)$ is in fact a function of the $n^{th}$ and $2n^{th}$ moment of the random variable $X$, given by

$$\left| \frac{\sigma_n}{\mu_n} \right| = \frac{1}{\sqrt{N_s}} \left| \frac{\langle X^{2n} \rangle}{\langle X^n \rangle^2} - 1 \right|^{1/2}. \tag{8}$$

In order to have a reliable measurement of $\langle X^n \rangle$, the PDF associated to $M_n$ must therefore sharply peak around $\mu_n$, that is to say $M_n$ must be contained within the narrowest interval centered on $\mu_n$. A measure of this property is precisely given by the ratio $|\sigma_n/\mu_n|$ found in equation (8). More specifically, $\mu_n$ is a reliable estimate of $\langle X^n \rangle$ when $|\sigma_n/\mu_n| \ll 1$. From (8), this can only be achieved if the number of independent samples $N_s$ is large enough to balance out the quantity $\sqrt{|\langle X^{2n} \rangle/\langle X^n \rangle^2 - 1|}$, which is unknown a priori.

Equation (8) clearly illustrates the underlying issue with statistical convergence: reducing the ratio $|\sigma_n/\mu_n|$ by one order of magnitude implies increasing the number of independent samples by two, which poses some obvious experimental challenges. In order to get a prediction for the required recording length, we followed Podesta et al.'s observation according to which an estimate for the quantity $\sqrt{|\langle X^{2n} \rangle/\langle X^n \rangle^2 - 1|}$ may be determined empirically by fitting a $N_s^{-1/2}$ power law to a plot made of the quantity $|\sigma_n/\mu_n|$, computed for different sample sizes $N_s$. Extrapolating the curve to any wanted value of $|\sigma_n/\mu_n|$ ultimately yields an estimate for the number of samples required to reach the aforementioned accuracy. In practice we aimed for $|\sigma_n/\mu_n| \leqslant 0.1$ (see figure 6.top).

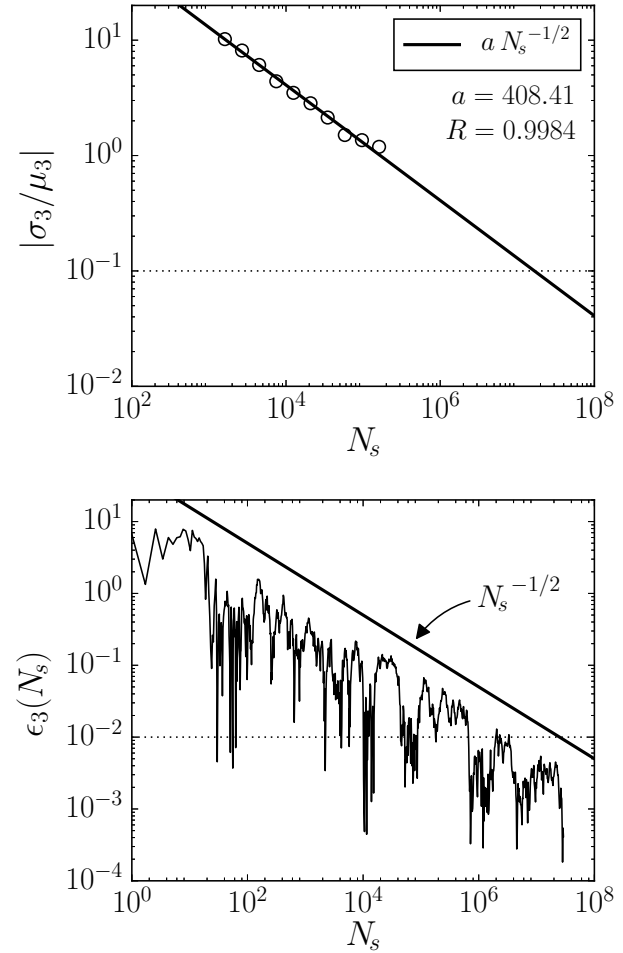

**Fig. 6** Statistical convergence of $\langle \|\delta \mathbf{u}'\|^2 \delta \mathbf{u}' \rangle$ for $r_\perp = 15\,\text{mm}$, $Ha = 10930$ ($B_0 = 3\,\text{T}$) and $Re^0 = 21320$ ($N_e = 64$, $I_0 = 160\,\text{A}$). Top: a priori estimation of $|\sigma_3/\mu_3|$ as a function of the number of samples. The calculation is based on an 18 mn long recording (ca. $2.1 \times 10^6$ samples). (······): $|\sigma_3/\mu_3| = 0.1$ threshold. $R$ represents the correlation coefficient of the curve fitting (in the least square sense). Bottom: a posteriori monitoring of the relative error $\epsilon_3$ based on 288 mn worth of data (ca. $3.4 \times 10^7$ samples). (······): 1% relative error mark achieved when $N_s \geqslant 10^7$.

Every time the control parameters were changed, an 18 mn long data set was recorded, on which we applied the procedure described above. We found that the ratio $|\sigma_n/\mu_n|$ for $n = 2$ and $n = 3$ seemed to depend neither on the magnitude of the magnetic field, nor on the intensity of the forcing. There however seemed to be a general trend according to which small velocity increments usually required one order of magnitude more samples than large velocity increments to reach the same level of accuracy. The procedure is illustrated in figure 6, which compares the a priori estimate for $N_s$, to the a posteriori monitoring of the convergence of $M_3$, associated to the random variable $\langle \|\delta \mathbf{u}'\|^2 \delta \mathbf{u}' \rangle$. The relative error associated to the latter was defined as

$$\epsilon_3 = \left| \frac{M_3(N_s) - M_3(N_{\max})}{M_3(N_{\max})} \right|, \qquad (9)$$

which compares $M_3(N_s)$ to its most accurate estimation computed using the maximum number of independent samples available $N_{\max}$. The scale chosen for this illustration was $r_\perp = 15$ mm. Furthermore, the operating settings were $Ha = 10930$ ($B_0 = 3$ T) and $Re^0 = 21320$ ($N_e = 64$, $I_0 = 160$ A), which are quite standard for the Flowcube, thus are representative overall. As figure 6 shows, one must roughly have $N_s \geqslant 10^7$ for $|\sigma_3/\mu_3|$ to be less than 0.1. Computing $M_3$ with $3.4 \times 10^7$ independent samples makes the relative error on the estimate to be less than 1%. This behavior was quite consistent throughout our data. Note also that the $N_s^{-1/2}$ decay of $\epsilon_3$ is in full agreement with equation (8).

By comparison, figure 7 shows the relative error for the convergence of the second order moment $M_2$ associated to the random variable $\langle \|\delta \mathbf{u}'\|^2 \rangle$. The convergence of $M_2$ is computed using the first $2 \times 10^6$ data points of the time series referred to earlier. Figure 7 illustrates why second order moments converge faster. Despite $\epsilon_2$ still following a $\sqrt{N_s}$ convergence rate, the relative error $\epsilon_2$ starts off at a much lower level than $\epsilon_3$. As a matter of fact, after averaging over the first hundred samples, $\epsilon_2$ is one order of magnitude lower than $\epsilon_3$. As a consequence, the 1% error mark is attained two orders of magnitude quicker for $\epsilon_2$ than for $\epsilon_3$ (more specifically $N_s \geqslant 10^5$ for the former vs. $N_s \geqslant 10^7$ for the latter).

## 6 Characterization of the turbulence driven in Flowcube

### 6.1 Flow topology

Figure 8 displays iso-contours of electric potential measured in the high probe density area located at the center of both top and bottom Hartmann plates (see figure 2). The area covered by these probes is $32.5 \times 32.5$ mm$^2$. The experimental configuration showcased in this figure consists of a flow forced by the $10 \times 10$ array of electrodes separated by $l_i = 5$mm. The electric forcing is kept constant at $Re^0 = 4780$ ($N_e = 100$, $I_0 = 56$ A), and the aspect of the flow is given for two different magnetic fields: $Ha = 3644$ and $36440$ ($B_0 = 1$ and 10 T respectively).

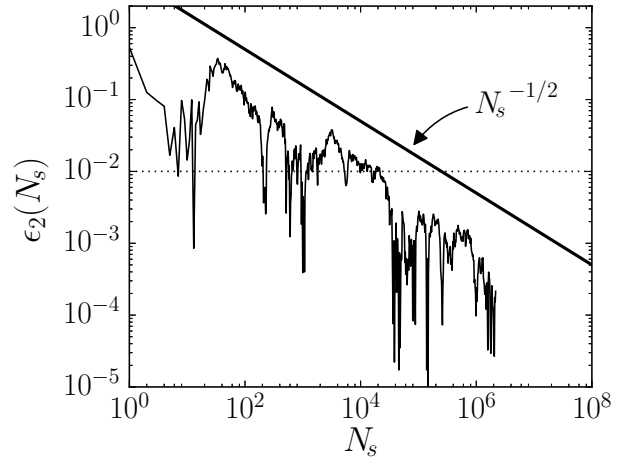

**Fig. 7** Relative error on the estimate for $\langle \|\delta \mathbf{u}'\|^2 \rangle$ for $r_\perp = 15$ mm, $Ha = 10930$ ($B_0 = 3$ T) and $Re^0 = 21320$ ($N_e = 64$, $I_0 = 160$ A), using an 18 mn long recording (ca. $2.1 \times 10^6$ samples); (······): 1% relative error mark achieved for $N_s \geqslant 10^5$.

Figure 8.left shows the iso-contours of mean electric potential $\bar{\Phi}(\mathbf{x}) = \langle \phi(\mathbf{x}, t) \rangle_t$, where $\langle \cdot \rangle_t$ is a time average performed at each point $\mathbf{x}$. Both top and bottom signals are normalized by $\bar{\Phi}_0 = \max(|\bar{\Phi}|)$ for each magnetic field respectively. The topology of the mean flow along the bottom wall (i.e. where the forcing takes place) is insensitive to the magnitude of the magnetic field, and consists of counter rotating vortices each centered on one electrode. In other words, the mean flow at this location is dominated by the topology of the forcing. As far as the mean flow along the top wall is concerned, the shape of the streamlines depends drastically on the value of the magnetic field. In particular, for a given electric forcing, increasing $B_0$ makes smaller and smaller structures become visible at the top of the experiment.

Figure 8.right shows a snapshot of electric potential fluctuations $\phi'(\mathbf{x}, t) = \phi(\mathbf{x}, t) - \bar{\Phi}(\mathbf{x})$ synchronously measured along the top and bottoms walls. Both top and bottom signals are normalized by $\phi'_0 = \max(|\phi'|)$ for each magnetic field respectively. The contours presented in figure 8.right may be seen as instantaneous pictures of the turbulent fluctuations generated in the experiment at a given instant. A short video showing the time evolution of electric potential fluctuations is available as supplementary material (the actual speed is twice as fast as the one used to compile the movie). Similarly to the mean flow studied earlier, small turbulent structures become more and more quasi-2D, as the magnetic field is increased. This behavior is evidenced qualitatively by noticing that the top and bottom pat-

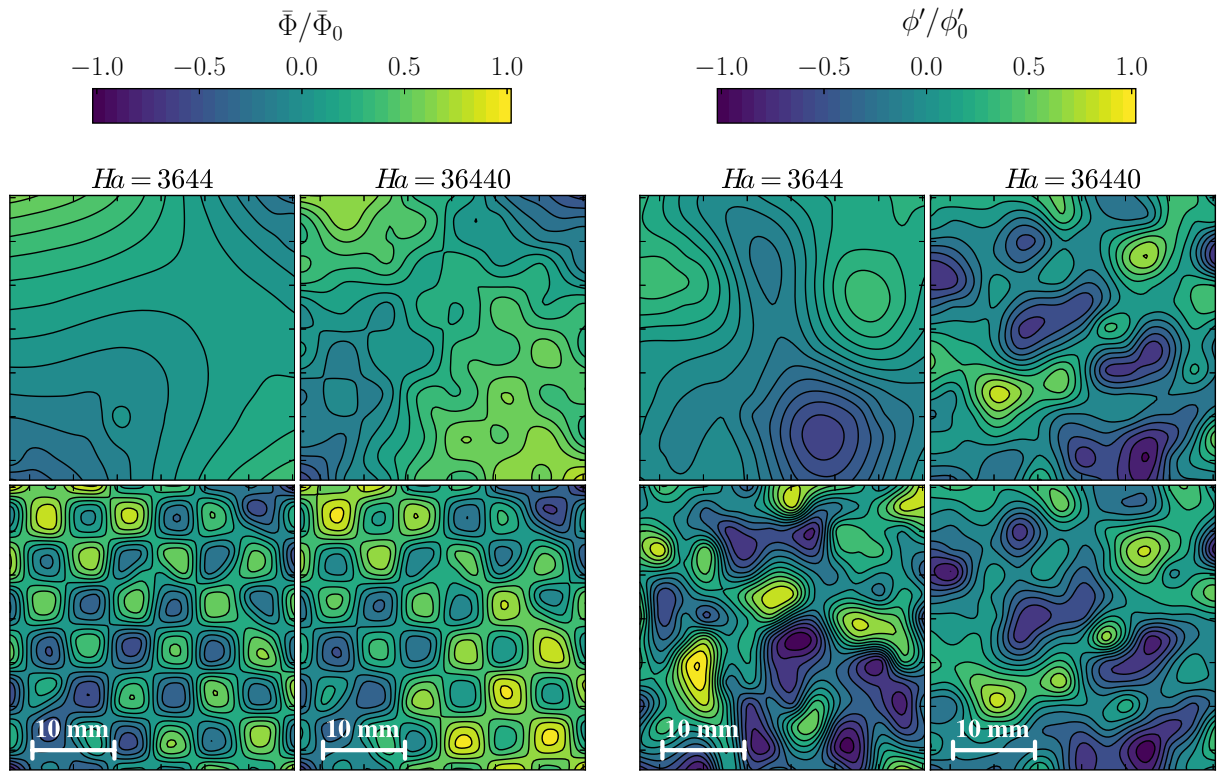

**Fig. 8** Iso-contours of the normalized mean electric potential $\bar{\Phi}/\bar{\Phi}_0$ (left) and electric potential fluctuations $\phi'/\phi'_0$ (right). The four top (resp. bottom) figures correspond to signals recorded along the top (resp. bottom) Hartmann wall. By definition, $\bar{\Phi}(\mathbf{x}) = \langle \phi(\mathbf{x},t) \rangle_t$ and $\phi'(\mathbf{x},t) = \phi(\mathbf{x},t) - \bar{\Phi}(\mathbf{x})$, while $\bar{\Phi}_0 = \max(|\bar{\Phi}|)$ and $\phi'_0 = \max(|\phi'|)$. Operating conditions consist of the $10 \times 10$ injection array with the injection scale $l_i = 5\,\mathrm{mm}$ and $Re^0 = 4780$ ($N_e = 100$, $I_0 = 56\,\mathrm{A}$). Two magnetic fields are given for comparison: $Ha = 3644$ and $36440$ ($B_0 = 1$ and $10\,\mathrm{T}$ respectively).

terns become increasingly mirror symmetrical at high magnetic fields.

These observations may be understood in the light of Sommeria & Moreau's [12] interpretation of the solenoidal component of the Lorentz force as a pseudo-diffusive process characterized by the diffusivity $\alpha \sim \sigma B_0^2 l_\perp^2 / \rho$, where $l_\perp$ is the width of a given turbulent scale. This argument shows that for a structure of a given width, increasing the magnetic field enables the Lorentz force to diffuse its momentum further and further in the direction of the magnetic field. This argument may also be seen the other way around: for a given magnetic field, the Lorentz force will diffuse the momentum of larger structures over a longer distance than smaller ones. As a result, increasing the magnetic field extends the range of 2D structures towards smaller scales.

Interestingly, the two-dimensionalizing effect of the Lorentz force seems to act differently on the base flow and the turbulent fluctuations. In particular, for $Ha = 36440$ (which is the most favorable settings to observe quasi-2D structures in this case), the turbulent fluctuations present a higher degree of top/bottom similarity than the mean flow. In other words, the dimensionality of turbulence is not necessarily dictated by the dimensionality of the forcing. This phenomenon was also observed by Pothérat and Klein [5].

## 6.2 PUDV / EPV benchmarking

Let us now compare the readings given by EPV and PUDV. Since these two methods of measurement are not available at the exact same location in the experiment, the following benchmarking was performed by considering the ultrasound probes closest to the top and bottom walls, and comparing them to the EPV signals measured along the top and bottom walls respectively. In all cases, the measurements were made along the same horizontal line running through the middle of the injection patch. The heights of the top and bottom ultrasound probes are $z = 88\,\mathrm{mm}$ and $z = 12\,\mathrm{mm}$ respectively. Since the Hartmann layers are extremely thin compared to the height of the channel, we will confuse their locations with those of the top and bottom walls located at $z = 100\,\mathrm{mm}$ and $z = 0\,\mathrm{mm}$ respectively.

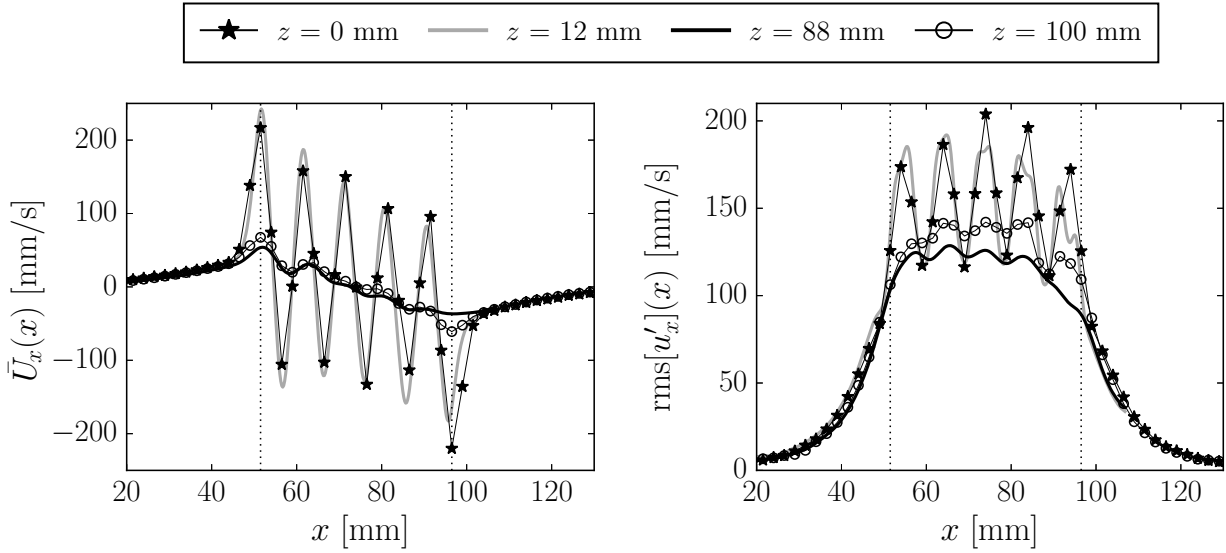

**Fig. 9** Electric Potential velocimetry (★, ○) vs. Pulsed Ultrasound velocimetry (▬, ▬) at the top and bottom of the cube; $Ha = 36440$ ($B_0 = 10$ T), $Re^0 = 17060$ ($N_e = 100$, $I_0 = 200$ A). Left: profile of the mean velocity component $\bar{U}_x(x) = \langle u_x(x,t)\rangle_t$. Right: profile of the turbulent fluctuations $\mathrm{rms}[u'_x]$. Vertical dotted lines demarcate the forcing area of the case at hand.

This convention was adopted for clarity, but should not hide the fact that the velocity at the walls are evidently null as a result of the no-slip boundary condition, and that EPV in fact gives a measure of the velocity field right outside the Hartmann layers. From now on, the EPV and PUDV signals measured at heights 100 mm and 88 mm respectively will be referred to as the top signals, while the EPV and PUDV signals measured at 0 mm and 12 mm respectively will be referred to as the bottom signals.

The operating parameters chosen for this benchmarking are $l_i = 5$ mm, $Ha = 36440$ ($B_0 = 10$ T), and $Re^0 = 17060$ ($N_e = 100$, $I_0 = 200$ A). These settings were chosen for two reasons. First of all, the smaller injection scale was used to highlight the superior spatial sampling rate of the PUDV method, equal to 0.8 mm in this particular case. Indeed, turbulent structures of size $l_i = 5$ mm lay at the bottom limit of EPV's detection range, whose spatial resolution of 2.5 mm is given by the distance separating two adjacent potential probes. Second of all, $Re^0 = 17060$, $Ha = 36440$ yields a reasonably quasi-2D turbulent flow with the smaller injection scale. As a matter of fact, the dimensionality of the forcing scale $l_i$ is typically of the order $l_z(l_i)/h \sim 1.9$ in this particular case, meaning that the turbulent statistics should be relatively invariant across $z$. As a result, the velocity profiles measured in the top or bottom portion of the experiment should be directly comparable, whether they stem from EPV or PUDV measurements.

Figure 9.left gives the spatial distribution of the mean velocity component $\bar{U}_x(x) = \langle u_x(x,t)\rangle_t$. The first obvious feature of this graph is the presence of five positive and negative peaks whose locations, and wavelength coincide with those of the electrodes in use. The presence of these peaks is therefore a marker of the electric forcing. It can be observed that the bottom, but also the top signals respectively compare with each other, and capture the same flow features. Figure 9.left shows that despite the extreme magnetic field, there is still some flagrant three dimensionality left in the mean flow, which can be visualized by comparing the top and bottom profiles and noticing that the amplitude of the former is 4 to 5 times weaker than that of the latter.

Figure 9.right shows the rms profile of the turbulent fluctuations $u'_x(x,t) = u_x - \bar{U}_x(x)$. As with the mean flow, the turbulent fluctuations captured by the EPV and PUDV method at the bottom compare very well, while slight differences in amplitude appear between the top signals. This behavior can certainly be put on the account of the flow not being fully quasi-2D.

EPV and PUDV appear to be reliable methods of measurement, which yield comparable results. More quantitatively, the relative discrepancy between EPV and PUDV readings can be assessed by introducing the quantities $e_{\mathrm{bot}}$ and $e_{\mathrm{top}}$ defined as

$$e_{\mathrm{bot}} = \frac{\langle |F_{\mathrm{EPV}}(x, z=0) - F_{\mathrm{PUDV}}(x, z=12)|\rangle_x}{\sqrt{\langle F_{\mathrm{EPV}}^2(x, z=0)\rangle_x}}, \quad (10)$$

and

$$e_{\text{top}} = \frac{\langle |F_{\text{EPV}}(x, z=100) - F_{\text{PUDV}}(x, z=88)| \rangle_x}{\sqrt{\langle F^2_{\text{EPV}}(x, z=100)\rangle_x}}, \quad (11)$$

where $F$ is a shorthand to designate profiles of $\bar{U}_x$ or rms$[u'_x]$, while the subscripts EPV and PUDV refer to the method of measurement used to get these profiles. Table 3 shows that in the conditions displayed here, the relative discrepancy between the two methods of measurement are in the 20% range on the mean flow, and 10% on the turbulent fluctuations. The larger discrepancy between the two methods of measurement observed when considering the mean flow comes from the high amplitude oscillations around zero.

**Table 3** Relative discrepancy between EPV and PUDV readings, when considering mean flow profiles ($F = \bar{U}_x$) and profiles of turbulent fluctuations ($F = \text{rms}[u'_x]$).

|  | $F = \bar{U}_x$ | $F = \text{rms}[u'_x]$ |
|---|---|---|
| $e_{\text{bot}}$ | 24% | 8% |
| $e_{\text{top}}$ | 23% | 10% |

As one can see, EPV and PUDV are complementary methods of measurement. On the one hand, PUDV can capture some of the smaller features of turbulence thanks to its higher spatial resolution. On the other hand, EPV can time resolve the flow thanks to its superior acquisition frequency. Although both systems may be used to compute the flow's statistical properties by assuming ergodicity, EPV was favored over PUDV for the computation of third order statistics. Indeed, the former could capture a much higher amount of statistical events within a given time-frame, thus reducing the time spent at a particular set point.

6.3 Dimensionality and componentality of the flow

We shall now illustrate how the dimensionality of the turbulence driven in the Flowcube is fully described by the diffusion lengthscale $l_z(l_i)$, associated to turbulent structures of the size of the forcing scale $l_i$. In the following, we characterize the dimensionality of the bulk by comparing the kinetic energy at height $z$ away from the forcing area $E_\perp(z) = \langle u'^2_x(x,z)\rangle_{t,x}$, to the kinetic energy injected into the flow. The latter is associated to the kinetic energy measured along the bottom Hartmann wall $E_\perp(0) = \langle \mathbf{u}'^2_\perp(\mathbf{x}_\perp)/2\rangle_{t,\text{bot}}$. Here, the operators $\langle \cdot \rangle_{t,x}$ and $\langle \cdot \rangle_{t,\text{bot}}$ represent time and space averaging over the horizontal beam of a transducer located at height $z$, and along the bottom wall respectively. The rationale for such a definition is the following: if turbulence is 2D over the distance $z$, then the energy content found at both ends must be the same, hence the ratio $E_\perp(z)/E_\perp(0)$ must be equal to one. Conversely, any kinetic energy deficit (thus embodied by a ratio $E_\perp(z)/E_\perp(0)$ lower than one) is a hint of velocity gradients existing over the distance $z$, hence of remaining three-dimensionality in the bulk.

In the spirit of Baker et al. [22], the ratio $E_\perp(z)/E_\perp(0)$ is plotted in figure 10 against the reduced height $z/l_z(l_i)$, where a global estimate for the diffusion length $l_z(l_i)$ is assessed in our experiment based on the rms of the horizontal turbulent fluctuations measured along the bottom plate (i.e. using EPV) $u'_{\text{bot}} = \sqrt{E_\perp(0)}$.

The collapse of all the data onto single curves proves that despite the flow originating from different operating conditions, the dimensionality of the flow is indeed fully contained within the diffusion length of turbulent structures of the size of the forcing scale. In particular, the dimensionality measured at a given distance $z$ away from the forcing only depends on how $z$ compares with the diffusion length $l_z(l_i)$. Note that figure 10 actually displays two curves, which are slightly shifted. Each curve corresponds in fact to a particular method of measurement. Had EPV and PUDV yielded the exact same data, it is more than likely that these two curves would have collapsed on top of each other.

Let us now illustrate the relationship between componentality and dimensionality. As mentioned earlier, the dimensionality of the flow is defined through the magnitude of the velocity gradients in the direction of the magnetic field estimated by the ratio $l_z(l_i)/h$. We also recall the physical interpretation of the aforementioned ratio: $l_z(l_i)/h \ll 1$ implies that turbulent structures of the size of the injection scale are topologically 3D (i.e. strong velocity gradients exist in the bulk), while $l_z(l_i)/h \gg 1$ means that the turbulent structures in question are quasi-2D (i.e. the flow is fully correlated along the direction parallel to the magnetic field).

Figure 11 shows the ratio of parallel to perpendicular turbulent kinetic energy (resp. $\bar{E}_\parallel$ and $\bar{E}_\perp$), computed from ultrasound signals. The former is defined as

$$\bar{E}_\parallel = \frac{1}{2}\left\langle \frac{1}{h}\int_0^h u'^2_\parallel(z,t)\,\mathrm{d}z \right\rangle_t, \quad (12)$$

where the integral is computed along the beam of a vertical transducer, and $u'_\parallel = \mathbf{u}' \cdot \mathbf{e}_z$ represents the turbulent velocity component aligned with the magnetic field. This definition makes it possible to estimate the level of parallel turbulent kinetic energy even in 3D cases, where the profile of energy is inhomogeneous along $z$.

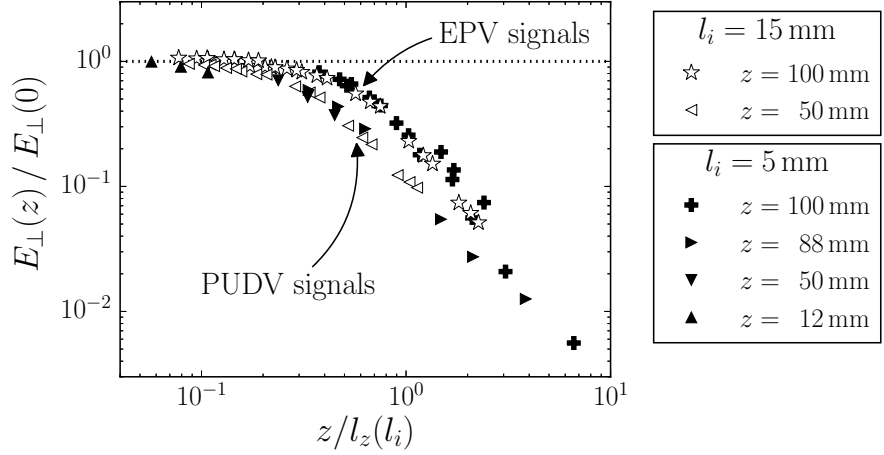

**Fig. 10** Dimensionality, measured by the ratio $E_\perp(z)/E_\perp(0)$ as a function of the parameter $z/l_z(l_i)$. Each marker corresponds to a distinct combination of operating parameters $(l_i, Ha, Re^0)$. $E_\perp(z=0\,\mathrm{mm})$ and $E_\perp(z=100\,\mathrm{mm})$ were obtained using EPV along the bottom and top Hartmann walls respectively, while $E_\perp(z=12,50,88\,\mathrm{mm})$ were calculated from PUDV measurements.

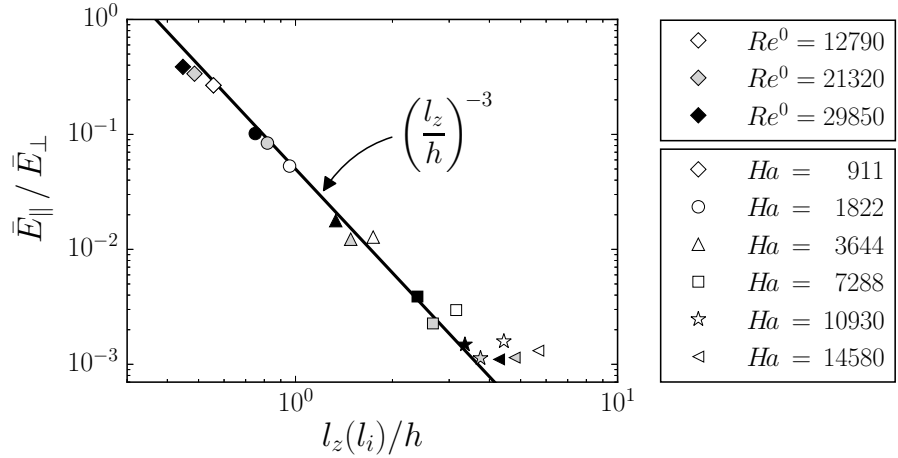

**Fig. 11** Componentality as a function of dimensionality. The ratio of vertical to horizontal kinetic energy is computed for different sets of magnetic fields $Ha$ and electric forcing $Re^0$, for the fixed forcing pattern consisting of the $8 \times 8$ injection array with the injection scale $l_i = 15\,\mathrm{mm}$.

Thanks to homogeneity in the horizontal plane, the perpendicular turbulent kinetic energy may simply be defined as a time and spatial average along the beam of the horizontal transducer located at mid height of the channel i.e. $\bar{E}_\perp = E_\perp(h/2)$.

Three-dimensional flows are characterized by values of $l_z(l_i)/h$ lower than unity. In particular, one can see from figure 11 that for $l_z(l_i)/h \simeq 0.4$, the vertical to horizontal energy ratio $\bar{E}_\parallel / \bar{E}_\perp$ is close to 0.5. This result is surprisingly close to what would be found in fully homogeneous and isotropic 3D turbulence. Despite this particular value of $\bar{E}_\parallel / \bar{E}_\perp$, we may not conclude that we are actually observing 3D homogeneous and isotropic turbulence, since the flow is inhomogeneous in the direction of the magnetic field. Nevertheless, this shows that when the bulk presents strong velocity gradients in the direction of the field, the vertical and horizontal velocity components are of the same order of magnitude. We are thus in presence of a three component velocity field. As the bulk becomes more and more two-dimensional (that is to say as $l_z(l_i)/h$ extends beyond unity), one can see that the amount of kinetic energy found in the vertical component becomes negligible compared to the horizontal one. As a matter of fact, the ratio plummets according to a steep $[l_z(l_i)/h]^{-3}$ law. The explanation for such a clear power law is unknown yet. It however suggests that there is indeed a link between dimensionality and componentality in our experiment, since points obtained using different operating conditions collapse onto the same curve.

Note that this behavior is not the sole consequence of the two-dimensionalization of the bulk by the Lorentz force, on the contrary. Indeed, Moffat [23] showed analytically that the two-dimensionalization of an unbounded flow by a magnetic field is in fact accompanied by the promotion of the vertical velocity component. In our case, one must see here the concurrent influence of the boundary conditions imposed by the impermeable horizontal walls, which forbid a vertical component in their vicinity (regardless of the dimensionality of the bulk). In fact, Pothérat & Kornet [24] showed numerically, within the context of decaying MHD turbulence between Hartmann walls, that walls indeed suppressed the velocity component aligned with the magnetic field.

# 7 Conclusion

This paper presented an experimental apparatus capable of driving wall bounded low-Rm MHD turbulence of controlled dimensionality. Thanks to the high probe density and number of electrodes introduced in this experiment, it is possible to drive and measure a wide range of turbulent scales, with a fine enough resolution. Thanks to the high acquisition rate of EPV, it is possible to compute statistics of second and third order which fall below the 1% convergence level, within a reasonable time frame. Owing to this high convergence level, we can assert that the Flowcube is indeed capable of reliably diagnosing the finest features of turbulence dynamics.

In addition, we showed that the kinematics of the flow driven in our experiment depended on the single non-dimensional parameter $l_z(l_i)/h$, which compares the range of action of the Lorentz force applying onto turbulent structures of the size of the forcing scale $l_i$ to the height of the channel. This succinct parameter quantifies the magnitude of the velocity gradients in the bulk. In a nutshell, $l_z(l_i)/h \ll 1$ implies that turbulent structures of size $l_i$ are 3D, while $l_z(l_i)/h \gg 1$ implies that structures of size $l_i$ extend throughout the channel, i.e. they are quasi-2D.

Finally, we showed that the presence of no-slip and impermeable walls perpendicular to the magnetic field introduces a strong link between the dimensionality of the flow as quantified by $l_z(l_i)/h$ and its componentality as measured by $\bar{E}_\parallel/\bar{E}_\perp$. In particular, wall-bounded low-Rm MHD turbulence appears to become two-component, as it becomes two-dimensional. This last point certainly shows that one must be cautious when interpreting results stemming from numerical simulations involving periodic boundary conditions.

**Acknowledgements** We wish to thank Ian Bates, Paul Chometon and Jurgen Spitznagel for their decisive contributions to designing and building the experimental apparatus.

We acknowledge financial support from the CNRS under the DEFI "instrumentation aux limites, VELOCIPEDE" scheme led by A. Sulpice.

A. Pothérat acknowledges support from the Royal Society under the *Wolfson Research Merit Award scheme* (grant reference WM140032) and the *International Exchanges scheme* (grant reference IE140127), from Université Grenoble-Alpes and from Grenoble-INP for the invited Professor positions they have granted him during the course of this project.

The laboratory SIMaP is part of the LabEx Tec 21 (Investissements d'Avenir - Grant Agreement No. ANR-11-LABX-0030).